\documentclass[11pt]{article}
\usepackage[left=1in,top=1in,right=1in,bottom=1in]{geometry}
\usepackage{times}

\usepackage{verbatim}
\usepackage{amsmath}
\usepackage{amssymb}
\usepackage{amsthm}
\usepackage{rotating}
\usepackage{algorithm}
\usepackage[noend]{algpseudocode}

\usepackage{silence}
\usepackage{tabularx}
\usepackage{graphicx}
\usepackage{url}
\usepackage{multirow}
\usepackage{subfig}

\usepackage[noend]{algpseudocode}
\usepackage{siunitx}

\DeclareMathOperator*{\argmax}{arg\,max}

\allowdisplaybreaks

\begin{document}
\title{Fictitious Play with Maximin Initialization}
\author{Sam Ganzfried\\
Ganzfried Research\\
sam.ganzfried@gmail.com
}

\date{\vspace{-5ex}}

\maketitle

\begin{abstract}
Fictitious play has recently emerged as the most accurate scalable algorithm for approximating Nash equilibrium strategies in multiplayer games. We show that the degree of equilibrium approximation error of fictitious play can be significantly reduced by carefully selecting the initial strategies. We present several new procedures for strategy initialization and compare them to the classic approach, which initializes all pure strategies to have equal probability. The best-performing approach, called maximin, solves a nonconvex quadratic program to compute initial strategies and results in a nearly 75\% reduction in approximation error compared to the classic approach when 5 initializations are used.
\end{abstract}

\section{Introduction}
\label{se:intro}
Nash equilibrium is the central solution concept in game theory. While a Nash equilibrium can be computed in polynomial time for two-player zero-sum games, it is PPAD-hard for two-player general-sum and multiplayer games and widely believed that no efficient algorithms exist~\cite{Chen05:Nash,Chen06:Settling,Daskalakis09:Complexity}. The best algorithm for computing an exact Nash equilibrium in multiplayer games is based on a non-convex quadratic program formulation and only scales to relatively small games~\cite{Ganzfried20:Fast}. For larger games several iterative algorithms have been considered; however, they have no theoretical guarantees and may have an extremely high degree of error. It has recently been shown that fictitious play produces a smaller degree of equilibrium approximation error in these games than regret minimization~\cite{Ganzfried20:Fictitious}, though the average error still becomes relatively large as the game size increases. For example, for 3-player games with 10 strategies per player and all payoffs uniform random in [0,1], the average equilibrium error from fictitious play is 0.056. The classic version of fictitious play initializes strategies for all players to play all actions with equal probability. In this paper we will explore more sophisticated initialization approaches to improve the algorithm's performance. 

A \emph{strategic-form game} consists of a finite set of players $N = \{1,\ldots,n\}$, a finite set of pure strategies $S_i$ for each player $i \in N$, and a real-valued utility for each player for each strategy vector (aka \emph{strategy profile}), $u_i : \times_i S_i \rightarrow \mathbb{R}$. A \emph{mixed strategy} $\sigma_i$ for player $i$ is a probability distribution over pure strategies, where $\sigma_i(s_{i'})$ is the probability that player $i$ plays pure strategy $s_{i'} \in S_i$ under $\sigma_i$. Let $\Sigma_i$ denote the full set of mixed strategies for player $i$. A strategy profile $\sigma^* = (\sigma^*_1,\ldots,\sigma^*_n)$ is a \emph{Nash equilibrium} if $u_i(\sigma^*_i,\sigma^*_{-i}) \geq u_i(\sigma_i, \sigma^*_{-i})$ for all $\sigma_i \in \Sigma_i$ for all $i \in N$, where $\sigma^*_{-i} \in \Sigma_{-i}$ denotes the vector of the components of strategy $\sigma^*$ for all players excluding player $i$. Here $u_i$ denotes the expected utility for player $i$, and $\Sigma_{-i}$ denotes the set of strategy profiles for all players excluding player $i$. It is well known that a Nash equilibrium exists in all finite games~\cite{Nash50:Non-cooperative}. In practice all that we can hope for is convergence of iterative algorithms to an approximation of Nash equilibrium. For a given candidate strategy profile $\sigma^*$, define $\epsilon(\sigma^*) = \max_i \max_{\sigma_i \in \Sigma_i} \left[ u_i(\sigma_i,\sigma^*_{-i}) - u_i(\sigma^*_i, \sigma^*_{-i}) \right]$. The goal is to compute a strategy profile $\sigma^*$ with as small a value of $\epsilon$ as possible (i.e., $\epsilon = 0$ would indicate that $\sigma^*$ comprises an exact Nash equilibrium). We say that a strategy profile $\sigma^*$ with value $\epsilon$ constitutes an \emph{$\epsilon$-equilibrium}. For two-player zero-sum games, there are algorithms with bounds on the value of $\epsilon$ as a function of the number of iterations and game size, and for different variations $\epsilon$ is proven to approach zero in the limit at different worst-case rates (e.g.,~\cite{Gilpin12:First}).

In classic fictitious play (Algorithm~\ref{al:classic-fp}), each player plays a best response to the average strategies of his opponents thus far~\cite{Brown51:Iterative,Robinson51:Iterative}. Strategies for all players can be initialized arbitrarily at $t = 0$; frequently they are initialized to play all pure strategies with equal probability. Then each player uses the following rule to obtain the average strategy at time $t$: 
$$\sigma^t _i = \left( 1 - \frac{1}{t+1} \right) \sigma^{t-1} _i + \frac{1}{t+1} \sigma'^t _i,$$
where $\sigma'^t _i$ is a best response of player $i$ to the profile $\sigma^{t-1} _{-i}$  of the other players played at time $t-1$.
Thus, the final strategy after $T$ iterations, $\sigma^T$, is the average of the strategies played in the individual iterations.

\begin{algorithm}
\caption{Classic fictitious play for $n$-player games}
\label{al:classic-fp}
\textbf{Inputs}: Game $G$, initial mixed strategies $\sigma^0_i$ for $i \in N$, number of iterations $T$
\begin{algorithmic}
\For {$t = 1$ to $T$}
\For {$i = 1$ to $n$}
\State $\sigma'^t_i = \argmax_{\sigma_i \in \Sigma_i} u_i(\sigma_i,\sigma^{t-1}_{-i})$ 
\State $\sigma^t_i = \left( 1 - \frac{1}{t+1} \right) \sigma^{t-1}_i + \frac{1}{t+1} \sigma'^t_i$ 
\EndFor
\EndFor
\Return $(\sigma^T_1,\ldots,\sigma^T_n)$
\end{algorithmic}
\end{algorithm}

For a game with $n$ players and $m$ actions per player, and $T$ iterations of fictitious play, Algorithm~\ref{al:classic-fp} runs in time $O(n m^n T).$ For each player we must compute a best response to strategy profile $\sigma^{t-1}_{-i}$ of the opponents. This requires iterating over all pure strategies for player $i$ and all joint strategy profiles for the opponents, of which there are $m^{n-1}.$ So the complexity of computing a best response for player $i$ is $O(m * m^{n-1}) = O(m^n).$ Since we must do this for $n$ players, the total complexity of the best response computations at each timestep is $O(n m^n).$ Note that the size of the game representation is $n m^n$, since we must represent a payoff for each player for each strategy profile. So we can view this procedure as being efficient despite the exponential dependence in the number of players (furthermore we are typically only interested in solving games for a small number of players). If we run the algorithm for $k$ different initializations, the complexity becomes $O(n m^n T k)$. We could parallelize the algorithm in various ways to improve speed if we have access to multiple cores. The most obvious way would be to run the different initializations on separate cores. We could also compute the best responses $\sigma'_i$ in parallel for the different players, as well as compute the expected value of each pure strategy in parallel for the best response calculation for a single player. In this paper we will be focusing on parallelization only over different initializations.

\section{Initialization approaches for fictitious play}
\label{se:init}
In this section we will describe several approaches for generating different initial strategy profiles to use for Algorithm~\ref{al:classic-fp}.
Once $K$ strategy profiles have been created, we can then run fictitious play in parallel using $K$ cores, and output the resulting strategies with smallest value of $\epsilon$ (Algorithm~\ref{al:fp-multi-init}). The most obvious approach for generating the strategies would be to set the probability of each pure strategy to be uniform in (0,1), then normalize~\ref{al:rand-naive}. However, it turns out that this approach does not actually generate a uniform-random strategy from the probability simplex for each player. In order to do this we must select the strategy probabilities from an exponential distribution, and normalize. This can be done straightforwardly using Algorithm~\ref{al:rand-init}. If we use the exponential distribution with parameter $\lambda$, the pdf is $f(x; \lambda) = e^{-\lambda x}$ and cdf is $F(x; \lambda) = 1-e^{-\lambda x}$, over the domain $x \geq 0.$ If we sample $U$ from Uniform(0,1) and set $T = F^{-1}(U)$, it turns out that $T$ has an exponential distribution,
where $F^{-1}$ is the quantile function, $$F^{-1}(p) = \frac{-\ln(1-p)}{\lambda}.$$
Furthermore if $U$ is uniform on (0,1) then $1-U$ is as well. So we can generate a sample from $f$ using
$T = \frac{-\ln U}{\lambda}$~\cite{Wiki22:Exponential}.
We can achieve our goal of generating a uniform-random strategy from the simplex by using any $\lambda > 0$, so we will just use $\lambda = 1$.

\begin{algorithm}
\caption{Fictitious play with multiple initializations}
\label{al:fp-multi-init}
\textbf{Inputs}: Game $G$, set of $K$ initial mixed strategies $\sigma^0_{k,i}$ for $i \in N$ $k = 1\ldots,K$, number of iterations $T$
\begin{algorithmic}
\State $\epsilon^* = \infty$
\For {$k = 1$ to $K$}
\For {$t = 1$ to $T$}
\For {$i = 1$ to $n$}
\State $\sigma'_{k,i} = \argmax_{\sigma_i \in \Sigma_i} u_i(\sigma_i,\sigma^{t-1}_{k,-i})$ 
\State $\sigma^t_{k,i} = \left( 1 - \frac{1}{t+1} \right) \sigma^{t-1}_{k,i} + \frac{1}{t+1} \sigma'^t_{k,i}$ 
\EndFor
\EndFor
\State $\epsilon_k = \max_i \max_{\sigma_i} \left[ u_i(\sigma_i,\sigma^T_{k,-i}) - u_i(\sigma^T_i, \sigma^T_{k,-i}) \right]$
\If {$\epsilon_k < \epsilon^*$}
\State $\sigma^* = \sigma^T_k$
\State $\epsilon^* = \epsilon_k$
\EndIf
\EndFor
\Return $\sigma^*$
\end{algorithmic}
\end{algorithm}

\begin{algorithm}
\caption{Na\"ive algorithm for generating initial strategies}
\label{al:rand-naive}
\textbf{Inputs}: Game $G$ with $n$ players and $m$ strategies per player
\begin{algorithmic}
\For {$i = 1$ to $n$}
\State $Z_i = 0$
\For {$j = 1$ to $m$}
\State $\sigma_i(j) = $ uniform random number in (0,1)
\State $Z_i = Z_i + \sigma_i(j)$
\EndFor
\For {$j = 1$ to $m$}
\State $\sigma_i(j) = \frac{\sigma_i(j)}{Z_i}$
\EndFor
\EndFor
\Return $\sigma = (\sigma_1,\ldots,\sigma_n)$
\end{algorithmic}
\end{algorithm}

\begin{algorithm}
\caption{Proper generation of uniform initial strategies}
\label{al:rand-init}
\textbf{Inputs}: Game $G$ with $n$ players and $m$ strategies per player
\begin{algorithmic}
\For {$i = 1$ to $n$}
\State $Z_i = 0$
\For {$j = 1$ to $m$}
\State $u = $ uniform random number in (0,1)
\State $\sigma_i(j) = -\ln (u)$
\State $Z_i = Z_i + \sigma_i(j)$
\EndFor
\For {$j = 1$ to $m$}
\State $\sigma_i(j) = \frac{\sigma_i(j)}{Z_i}$
\EndFor
\EndFor
\Return $\sigma = (\sigma_1,\ldots,\sigma_n)$
\end{algorithmic}
\end{algorithm}

One potential drawback of generating $K$ strategies per player uniformly at random is that we may happen to select strategies that are very similar,
eliminating the benefit of using multiple initializations. A similar problem has been observed with the $K$-means clustering algorithm. The standard version of $K$-means clustering selects the $K$ initial cluster centers to be random points from the dataset (aka ``MacQueen's method'')~\cite{MacQueen67:Some}. A second approach, called the maximin method, chooses the first center arbitrarily, and for each subsequent center selects the point that has the greatest minimum distance to a previously selected center~\cite{Gonzalez85:Clustering,Katsavounidis94:New}. The $K$-means++ method essentially interpolates between MacQueen's method and the maximin method by selecting each point to be the next center with probability proportional to its squared minimum distance from a previously selected center\cite{Arthur07:K-means}. In a comparative study it has been shown that MacQueen's method and the maximin method ``often perform poorly'' and are significantly outperformed by $K$-means++~\cite{Celebi13:Comparative}.

Note that the goal of determining good initializations for equilibrium computation with fictitious play is not necessarily the same as that of finding good initial centers for $K$-means clustering. While for clustering we would typically want centers that are fairly evenly spread throughout the space, it is not clear that this is best for fictitious play; for example, we may obtain better performance by choosing corner points or points at the edge of the strategy simplex and no points near the center. Furthermore, for $K$-means we are given an initial finite set of datapoints that are the candidate centers, while for fictitious play the initial points can be any points in the (infinite) strategy simplex. It is also not clear if randomness is helpful for selection of the initial strategies in fictitious play. 

If we wish to use approaches similar to the maximin method and $K$-means++ for fictitious play initialization, we have two options: we could apply them to the full space of strategy profiles, or we could first create a ``dataset'' of $H$ initial ``points'' selected uniformly at random (using Algorithm~\ref{al:rand-init}) and then apply the algorithms to this dataset. We can implement the maximin method on the full simplex by solving a non-convex optimization problem, though it seems computationally intractable for $K$-means++.  For the sampling approach, we can implement maximin initialization using Algorithm~\ref{al:maximin}, and $K$-means++ initialization using Algorithm~\ref{al:FP++}, which we refer to as FictitiousPlay++. For the distance metric between strategy profiles $\sigma = (\sigma_1,\ldots,\sigma_n)$ and $\sigma' = (\sigma'_1,\ldots,\sigma'_n)$ we use L2:
$$D(\sigma,\sigma') = \sqrt{\sum_{i = 1}^n\sum_{j = 1}^{m(=|S_i|)}\left(\sigma_i(j) - \sigma'_i(j)\right)^2}.$$

\begin{algorithm}
\caption{Maximin initialization for fictitious play}
\label{al:maximin}
\textbf{Inputs}: Game $G$ with $n$ players and $m$ strategies per player, number of initial strategy profiles $K$,
total number of sampled strategy profiles $H$
\begin{algorithmic}
\For {$h = 1$ to $H$}
\State $\sigma^h =$ strategy profile generated according to Algorithm~\ref{al:rand-init}
\EndFor
\State $\tau^1 =$ one of the strategy profiles $\sigma^h$ chosen uniformly at random
\For {$k = 2$ to $K$}
\For {each $\sigma^h$ not chosen as one of the $\tau^i$ yet}
\State Compute $D(\sigma^h)$, the distance between $\sigma^h$ and the nearest $\tau^i$ that has already been chosen
\State Choose $\tau^k$ to be point $\sigma^h$ with largest value of $D(\sigma^h)$
\EndFor
\EndFor
\Return $(\tau^1,\ldots,\tau^K)$
\end{algorithmic}
\end{algorithm}

\begin{algorithm}
\caption{FictitiousPlay++}
\label{al:FP++}
\textbf{Inputs}: Game $G$ with $n$ players and $m$ strategies per player, number of initial strategy profiles $K$,
total number of sampled strategy profiles $H$
\begin{algorithmic}
\For {$h = 1$ to $H$}
\State $\sigma^h =$ strategy profile generated according to Algorithm~\ref{al:rand-init}
\EndFor
\State $\tau^1 =$ one of the strategy profiles $\sigma^h$ chosen uniformly at random
\For {$k = 2$ to $K$}
\For {each $\sigma^h$ not chosen as one of the $\tau^i$ yet}
\State Compute $D(\sigma^h)$, the distance between $\sigma^h$ and the nearest $\tau^i$ that has already been chosen
\State Choose $\tau^k$ at random using a weighted distribution where point $\sigma^h$ is chosen with probability proportional to $D(\sigma^h)^2$ 
\EndFor
\EndFor
\Return $(\tau^1,\ldots,\tau^K)$
\end{algorithmic}
\end{algorithm}

We can implement the maximin method on the full simplex by solving the following quadratically-constrained program, which is nonconvex. The variable $x_{ij}$ denotes the probability that player $i$ plays pure strategy $j$, for $1 \leq i \leq n, 1 \leq j \leq m$. And the constant $\tau^i_{jk}$ is the probability that player $j$ plays pure strategy $k$ under the $i$th center for $0 \leq i \leq t$, where $t$ is the number of centers that have already been selected. The initial center $\tau^0$ is selected randomly according to Algorithm~\ref{al:rand-init}, and we solve the following program for $t = 1,2,\ldots,T.$ This program can be solved using Gurobi's nonconvex quadratic solver~\cite{Gurobi22:Gurobi}. Note that this approach will often result in adding extreme points (i.e., pure strategy profiles) for the initial values of $t.$

\small
\[
\begin{array}{rrl} 
&\max_{\mathbf{x}, y} & y \\
&\mbox{s.t.}& y \leq \sum_j \sum_k \left(\left(x_{jk}\right)^2 - 2\tau^i_{jk} x_{jk} + \left(\tau^i_{jk}\right)^2\right) \forall i\\
& & 0 \leq x_{ij} \leq 1 \forall i,j\\
& & \sum_{j} x_{ij} = 1 \forall i\\
\end{array} 
\]
\normalsize

In addition to using analogues of $K$-means initialization procedures for fictitious play initialization, we can also consider actually running $K$-means on a set of $H$ sampled points and outputting the computed cluster means. This would produce $K$ strategy profiles that are evenly spread throughout the simplex, and would be very unlikely to produce any near-extreme points. 

\section{Experiments}
\label{se:experiments}
In this section we experimentally evaluate the approaches described in the previous section. For MacQueen's method, we consider both the version that uses the correct uniform sampling (Algorithm~\ref{al:rand-init}) as well as a version using the na\"ive (and incorrect) sampling approach (Algorithm~\ref{al:rand-naive}). For the maximin method we consider both the approaches where the full strategy spaces are used and where the centers are selected from a set of sampled points from the simplex. For FictitiousPlay++ we only consider the sampled version, since the unsampled version is computationally intractable. We similarly only consider the sampled version for the method based on $K$-means.

\begin{enumerate}
\item Classic fictitious play (classic)
\item MacQueen's method using Algorithm~\ref{al:rand-naive} (macqueen-1)
\item MacQueen's method using Algorithm~\ref{al:rand-init} (macqueen-2)
\item Unsampled maximin initialization (maximin-u)
\item Sampled maximin initialization (maximin-s)
\item Sampled $K$-means++ initialization (fp++)
\item Initialization from $K$-means cluster centers (k-means)
\end{enumerate}

For our first set of experiments we compared the seven approaches on games with $n=3$ players and $m = 5$ pure strategies per player, with all payoffs uniformly random in [0,1]. We ran all of the algorithms on each of 10,000 randomly generated games. For each game, we computed five different initializations with each algorithm (except for classic where we just used one), and chose strategies from the initialization that produced the lowest $\epsilon$ out of the five (i.e., $\epsilon^*$ in Algorithm~\ref{al:fp-multi-init}). For each initialization, we ran fictitious play for 10,000 iterations, as this value had been established previously as a balance between running time and performance~\cite{Ganzfried20b:Parallel,Ganzfried20:Fictitious,Ganzfried21:Computing}.

For each random game we generated 100,000 random strategies (using Algorithm~\ref{al:rand-init}) to be used for the strategies requiring simulation (maximin-s, fp++, and k-means).  For each application of $K$-means we used the $K$-means++ initialization procedure, selecting the run that produced the clustering with lowest error out of five different $K$-means++ initializations. For each of these initializations, we ran $K$-means for 50 iterations (or until the clusters stopped changing). We use the $K = 5$ cluster centers from the best of the runs as the five initializations for fictitious play. 

For these experiments we used a server with 64 cores, though parallelization was only used for maximin-u, fp++, and $K$-means. For maximin-u the parallelization was used by Gurobi's nonconvex quadratic solver, while for fp++ and $K$-means the parallelization was used to iterate over the 100,000 total datapoints in parallel. In addition to parallelizing each step of $K$-means, we also used a pruning technique that exploits the triangle inequality~\cite{Elkan03:Using} to reduce the runtime. Note that these experiments could have run significantly faster had we not sampled new strategies or run $K$-means for each new game; however, doing these would make the samples of $\epsilon$ not be independent for all algorithms, preventing us from concluding statistical significance. 

The results from these experiments are in Table~\ref{ta:results}. The classic version of fictitious play produced the largest value of $\epsilon^*$, with all other algorithms obtaining significantly smaller $\epsilon^*.$ Interestingly $K$-means produced the worst results out of the remaining algorithms. This indicates that we do not actually want the initial strategies to be ``evenly spread'' throughout the space as cluster means, and that it is preferable to ensure that near-extreme points are included. The results indicate that macqueen-2 outperforms macqueen-1, as expected (though the performance difference is not that large). The results also indicate that both macqueen algorithms are outperformed by fp++, which in turn is outperformed by both maximin algorithms. 

\begin{table}[!ht]
\centering
\begin{tabular}{|*{2}{c|}} \hline
Algorithm & Average value of $\epsilon^*$ \\ \hline
classic &0.02213 $\pm$ \num{6.87e-4}  \\ \hline
macqueen-1 & 0.00750 $\pm$ \num{3.25e-4} \\ \hline
macqueen-2 & 0.00690 $\pm$ \num{3.10e-4} \\ \hline
maximin-u &0.00579 $\pm$ \num{2.77e-4} \\ \hline
maximin-s &0.00623 $\pm$ \num{2.93e-4} \\ \hline
fp++ &0.00686 $\pm$ \num{3.06e-4} \\ \hline
k-means &0.00818 $\pm$ \num{3.46e-4} \\ \hline
\end{tabular}
\caption{Results for uniform-random games with payoffs in [0,1] for $n = 3$, $m = 5.$ Results over 10,000 random games, with 5 initializations for each algorithm (other than classic) and 10,000 iterations of fictitious play per initialization. The lowest $\epsilon$ out of the 5 initializations was used. For the sampling algorithms we randomly generated 100,000 strategy profiles for each game. For each algorithm we report the average $\epsilon^*$ and the 95\% confidence interval.}
\label{ta:results}
\end{table}

Since several of the algorithms obtained statistically indistinguishable performance and $K$-means took longer to run than the other algorithms (while performing clearly worse), we next experimented on all the algorithms except for $K$-means on 100,000 games. The results in Table~\ref{ta:results-nokmeans} make it more clear that maximin outperforms the other approaches; in particular, maximin-u produces a nearly 75\% reduction in Nash approximation error. The results also indicate that the performances of fp++ and macqueen-2 are very similar. 

\begin{table}[!ht]
\centering
\begin{tabular}{|*{2}{c|}} \hline
Algorithm & Average value of $\epsilon^*$ \\ \hline
classic &0.02234 $\pm$ \num{2.18e-4}  \\ \hline
macqueen-1 & 0.00743 $\pm$ \num{1.02e-4} \\ \hline
macqueen-2 & 0.00685 $\pm$ \num{9.72e-5} \\ \hline
maximin-u &0.00603 $\pm$ \num{9.05e-5} \\ \hline
maximin-s &0.00622 $\pm$ \num{9.21e-5} \\ \hline
fp++ &0.00680 $\pm$ \num{9.65e-5} \\ \hline
\end{tabular}
\caption{Results for uniform-random games with payoffs in [0,1] for $n = 3$, $m = 5.$ Results over 100,000 random games, with 5 initializations for each algorithm (other than classic) and 10,000 iterations of fictitious play per initialization.}
\label{ta:results-nokmeans}
\end{table}

We next experimented with the classic, macqueen-2, maximin-u, and fp++ approaches on 10,000 games with $n = 3$, $m = 10$, with all the other parameters the same as before. The order of performance of the algorithms was the same as for $m = 5$, though macqueen-2, maximin-u, and fp++ all achieved very similar average $\epsilon^*$ and their performances were not statistically distinguishable. All three approaches produced a significant improvement over classic, reducing the average value of $\epsilon^*$ by nearly 75\%.

\begin{table}[!ht]
\centering
\begin{tabular}{|*{2}{c|}} \hline
Algorithm & Average value of $\epsilon^*$ \\ \hline
classic &0.05582 $\pm$ \num{8.94e-4}  \\ \hline
macqueen-2 & 0.01519 $\pm$ \num{4.18e-4} \\ \hline
maximin-u &0.01459 $\pm$ \num{4.11e-4} \\ \hline
fp++ &0.01467 $\pm$ \num{4.10e-4} \\ \hline
\end{tabular}
\caption{Results for uniform-random games with payoffs in [0,1] for $n = 3$, $m = 10.$ Results over 10,000 random games, with 5 initializations for each algorithm (other than classic) and 10,000 iterations of fictitious play per initialization. The lowest $\epsilon$ out of the 5 initializations was used. For fp++ we randomly generated 100,000 strategy profiles for each game.}
\label{ta:results-10actions}
\end{table}

We experimented with the same 4 algorithms on 10,000 games with $n = 4$, $m = 3$ and 5, using the same parameters as the prior experiments. 
The relative ordering of the algorithms' performances was the same as in the 3-player experiments, with the three new algorithms significantly
outperforming classic. The best-performing algorithm was maximin-u, reducing average $\epsilon^*$ by 70\% compared to classic.

\begin{table}[!ht]
\centering
\begin{tabular}{|*{3}{c|}} \hline
Algorithm &Avg. $\epsilon^*$ for $m = 3$ & Avg. $\epsilon^*$ for $m = 5$ \\ \hline
classic &0.01993 $\pm$ \num{6.00e-4} &0.04777 $\pm$ \num{7.55e-4} \\ \hline
macqueen-2 & 0.00724 $\pm$ \num{2.98e-4} &0.01511 $\pm$ \num{3.75e-4} \\ \hline
maximin-u &0.00650 $\pm$ \num{2.83e-4} &0.01422 $\pm$ \num{3.59e-4}\\ \hline
fp++ &0.00719 $\pm$ \num{3.00e-4} &0.01488 $\pm$ \num{3.69e-4}\\ \hline
\end{tabular}
\caption{Results for uniform-random games with payoffs in [0,1] for $n = 4$. Results over 10,000 random games for each value of $m$.}
\label{ta:results-4pl-5act}
\end{table}

\begin{figure}[!ht]
\centering
\includegraphics[scale=0.6]{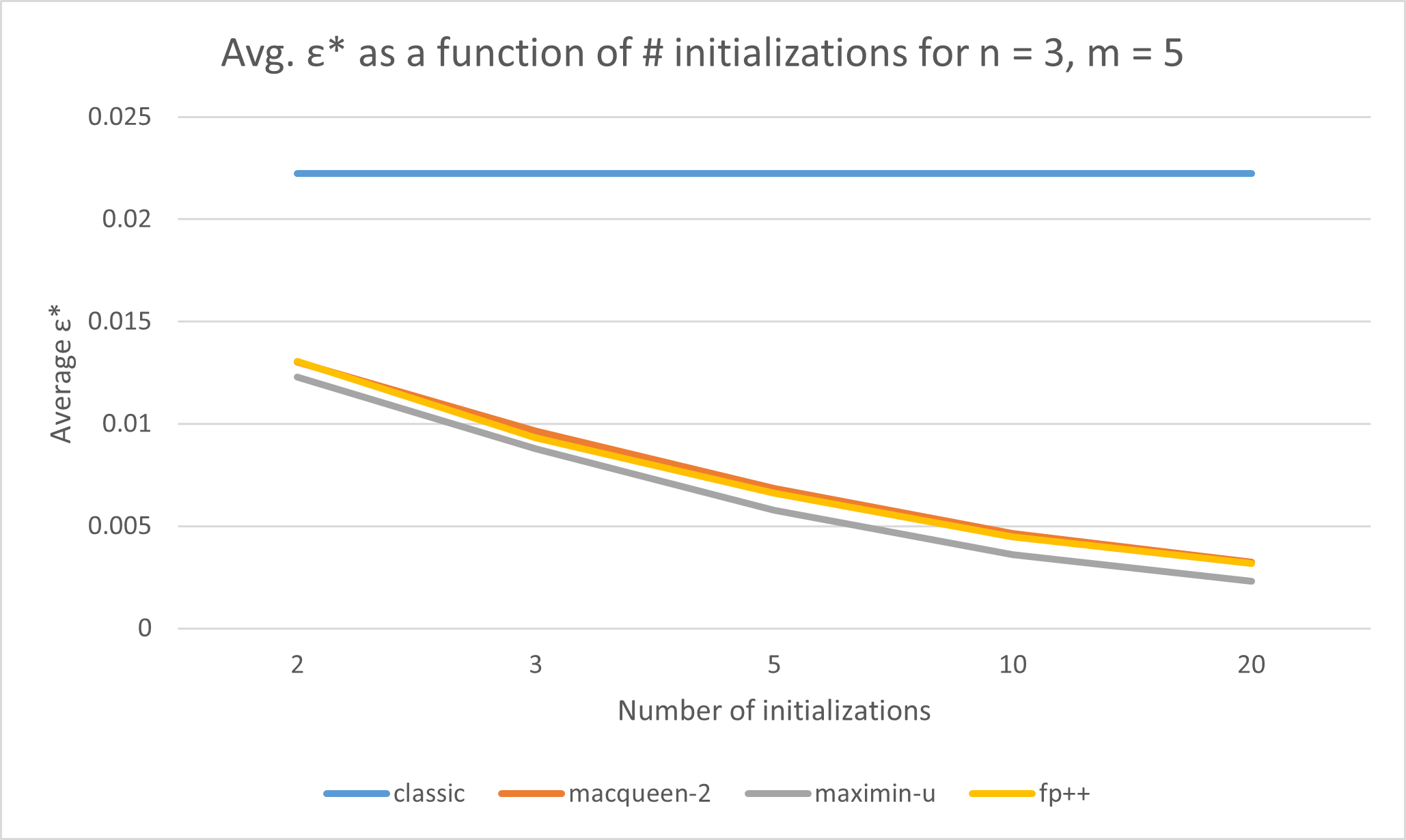}
\caption{Average $\epsilon^*$ as a function of number of initializations for $n = 3$, $m = 5.$ Results over 10,000 random games, with 2, 3, 5, 10, and 20 initializations for the algorithms.}
\label{fi:init-3-5}
\end{figure}

Next we explored the performances of the algorithms for different numbers of initializations $K$ (in all the previous experiments we have used $K = 5$). We varied $K$ to be 2, 3, 5, 10, and 20 for several values of $m$ and $n$. For each setting of $m$ and $n$, we generated 10,000 uniform random games with payoffs in [0,1], using the same values as before for the other parameters. Our first experiments were for $n = 3$ and $m = 5.$ As before the relative ordering of the algorithms by performance was maximin-u, fp++, then macqueen-2. From Figure~\ref{fi:init-3-5} we can see that just using $K = 2$ resulted in a performance improvement of 45\% for maximin-u over classic, and using $K = 20$ resulted in an improvement of 90\% (nearly a full order of magnitude). For these experiments we divided up the 10,000 games over 64 cores and just used a single core for each algorithm.

\begin{figure}[!ht]
\centering
\includegraphics[scale=0.65]{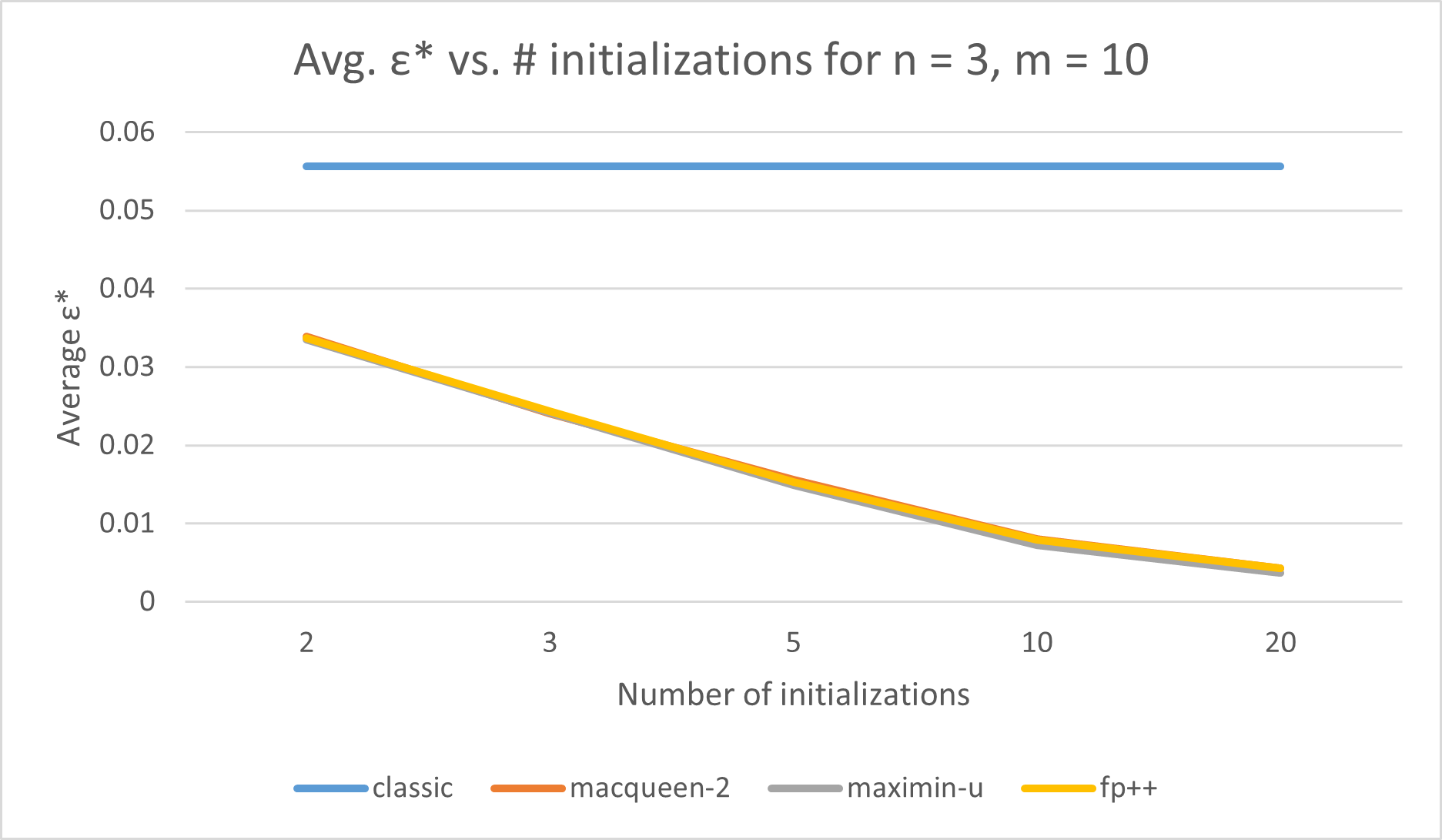}
\caption{Average $\epsilon^*$ as a function of number of initializations for $n = 3$, $m = 10.$ Results over 10,000 random games, with 2, 3, 5, 10, and 20 initializations for the algorithms.}
\label{fi:init-3-10}
\end{figure}

\begin{figure}[!ht]
\centering
\includegraphics[scale=0.65]{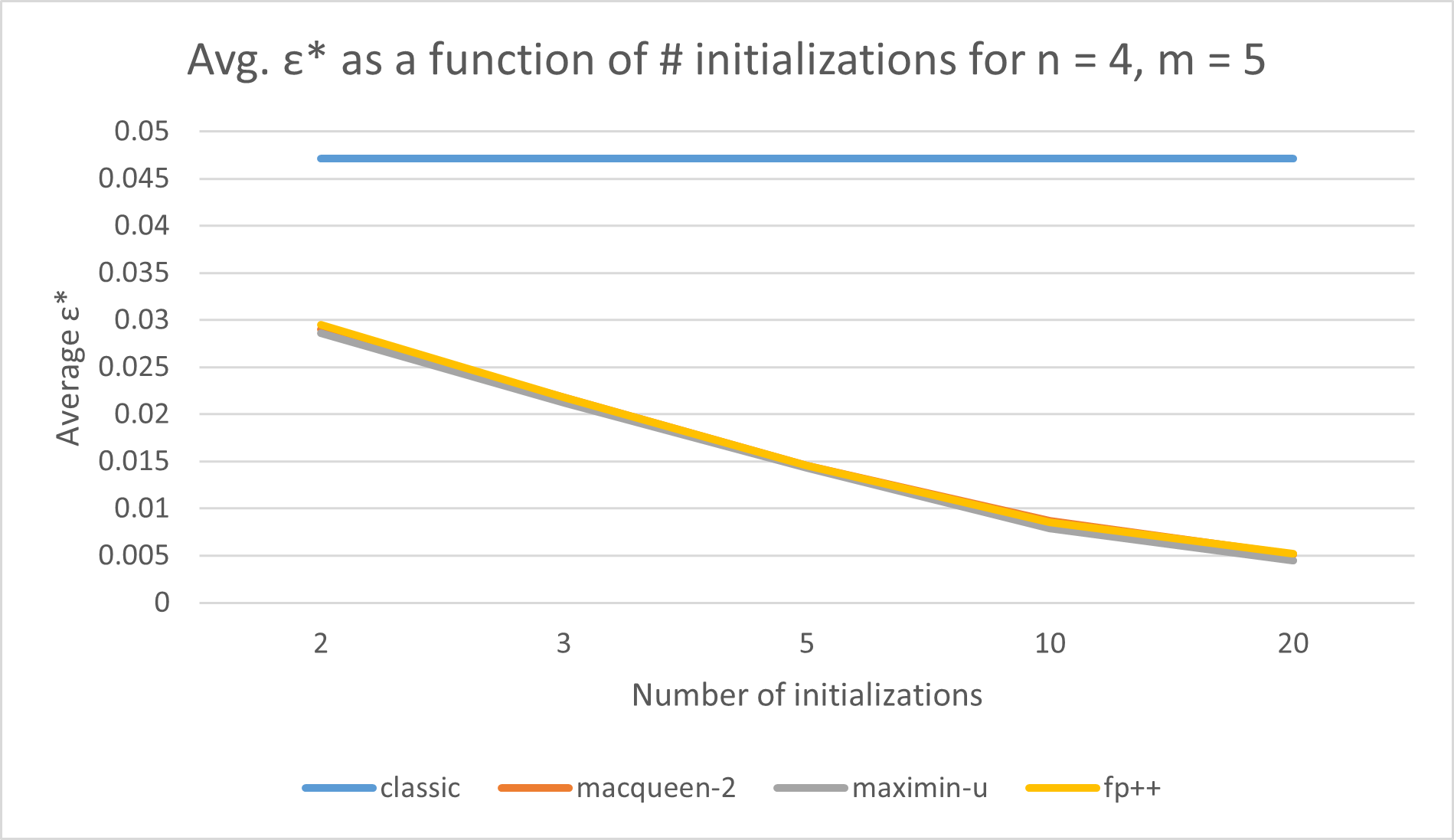}
\caption{Average $\epsilon^*$ as a function of number of initializations for $n = 4$, $m = 5.$ Results over 10,000 random games, with 2, 3, 5, 10, and 20 initializations for the algorithms.}
\label{fi:init-4-5}
\end{figure}

Results for similar experiments using $n = 3$, $m = 10$ are provided in Figure~\ref{fi:init-3-10}. The relative performance ordering of the algorithms remains the same, and the improvement over classic is more significant than for $m = 5.$ Using $K = 20$ maximin-u reduces average $\epsilon^*$ by 93\% compared to classic. We also performed similar experiments with $n = 4$, $m = 5$ (Figure~\ref{fi:init-4-5}). Again maximin-u outperformed the other algorithms, reducing average $\epsilon^*$ by 90\% for $K = 20.$ 

\section{Discussion}
\label{se:disc}
We have seen that we can obtain significantly smaller Nash equilibrium approximation error over classic fictitious play by selecting the best run over multiple initializations, even for a small number of initializations. We often achieved a performance improvement over 40\% just using $K = 2$, which gradually increased for larger $K$ and often achieved a 90\% improvement for $K = 20.$ We have seen a consistent ordering of the algorithms' performance which was statistically significant in many of the experiments: maximin followed by fp++ followed by macqueen followed by k-means. The results not only indicate that using multiple initializations is important, but furthermore that it is beneficial to ensure that near-extreme points are used. The worst-performing algorithm was standard $K$-means, which produces cluster centers that are relatively evenly spread throughout the joint strategy space, making it very unlikely to select any near-extreme points. It is interesting that $K$-means is outperformed by macqueen, which samples points uniformly and runs the risk of selecting several initial strategies that are close together. This indicates that it is more important to ensure that near-extreme points are included than it is to ensure that the initial points are spread out evenly.

It is interesting to compare the performance of the algorithms for fictitious play initialization to analogues for $K$-means initialization. 
For $K$-means, where the ultimate goal is to produce cluster means that are evenly spread throughout the space, it has been shown that MacQueen's method and the maximin method perform poorly and are significantly outperformed by $K$-means++~\cite{Celebi13:Comparative}. For fictitious play, we have shown that the maximin method outperforms the others. Note that all three of these approaches achieve a significant improvement over the classic approach, and the difference in performance between them was typically quite small. While maximin consistently performed the best in these experiments, it is possible that macqueen or fp++ can perform better in game classes other than uniform random; in particular, for games whose Nash equilibria involve a large amount of randomization.

For several of the algorithms we had to first sample a large number of points from the joint strategy space (maximin-s, fp++, and k-means). In order to obtain independence in the experiments, we sampled a new set of points for each game (and furthermore we used the same set of sampled points for all algorithms to reduce variance). We also came up with a new quadratic program formulation for implementing the maximin approach without requiring these samples, resulting in algorithm maximin-u which outperformed the sampled version maximin-s. We also compared MacQueen's method using the correct method for generating random points from the simplex (macqueen-2) to a simple yet incorrect approach (macqueen-1). Not surprisingly macqueen-2 outperformed macqueen-1, though macqueen-1 still obtained a significant improvement over the classic algorithm. 

\section{Conclusion}
\label{se:conc}
Nash equilibrium is the central solution concept in game theory, and computing (or approximating) one in games with more than two players is a notoriously challenging problem~\cite{Chen05:Nash,Chen06:Settling,Daskalakis09:Complexity}. No scalable algorithms exist with theoretical guarantees on performance. The best algorithm for computing an exact Nash equilibrium in multiplayer games is based on a non-convex quadratic program formulation and only scales to relatively small games~\cite{Ganzfried20:Fast}. For larger games several iterative algorithms have been considered; however, they have no theoretical guarantees and may have an extremely high degree of error. It has recently been shown that fictitious play produces a smaller degree of equilibrium approximation error in these games than another scalable algorithm called counterfactual regret minimization (CFR)~\cite{Ganzfried20:Fictitious}, though the average error still becomes relatively large as the game size increases. For example, for 3-player games with 10 strategies per player and all payoffs uniform random in [0,1], the average error of fictitious play is 0.056. CFR~\cite{Zinkevich07:Regret} has produced strategies for six-player no-limit Texas hold 'em that defeated strong human professionals~\cite{Brown19:Superhuman}; however there are no guarantees on its convergence to equilibrium, and it was shown to not converge to equilibrium in the  simplified game of three-player Leduc hold 'em~\cite{Abou10:Using}. While both fictitious play and CFR are guaranteed to converge to Nash equilibrium in two-player zero-sum games, they have no general performance guarantees in non-zero-sum and multiplayer games.

The classic version of fictitious play initializes strategies for all players to play all actions with equal probability. We have developed approaches that use multiple initializations to achieve a significant reduction in Nash equilibrium approximation error over the classic version, in some cases by a full order of magnitude. The best-performing algorithm, maximin, computes the initial strategies by solving a nonconvex quadratically-constrained program. Using this approach, we can compute close approximations of Nash equilibrium strategies in large multiplayer games for the first time.

\bibliographystyle{plain}
\bibliography{C://FromBackup/Research/refs/dairefs}

\end{document}